\pdfoutput=1

\documentclass[]{spie}  
\usepackage[]{graphicx}
\usepackage{xcolor}

\usepackage{fancyhdr} 
\fancypagestyle{plain}{%
    \fancyhf{}%
    \fancyfoot[C]{\thepage}%
}

\title{Quantifying telescope phase discontinuities external to AO-systems by use of Phase Diversity and Focal Plane Sharpening} 
\author{Masen P. Lamb\supit{a,b}, Carlos Correia\supit{c}, Jean-Fran\c{c}ois Sauvage\supit{c,d}, Jean-Pierre V\'{e}ran\supit{b}, David R. Andersen\supit{b}, Arthur Vigan\supit{c}, Peter L. Wizinowich\supit{e}, Marcos A. van Dam\supit{f}, Laurent Mugnier\supit{d}, Charlotte Bond\supit{c} 
\skiplinehalf
\supit{a}University of Victoria, 3800 Finnerty Rd, Victoria, Canada; \\
\supit{b}NRC Herzberg Astronomy, 5071 W. Saanich Rd, Victoria, Canada; \\
\supit{c}Aix Marseille Universit\'{e}, CNRS, LAM (Laboratoire d'Astrophysique de Marseille) UMR 7326, 13388, Marseille, France;\\
\supit{d}ONERA, 29 Avenue de la Division Leclerc, 92320 Ch\^{a}tillon, France; \\
\supit{e}W. M. Keck Observatory, 65-1120 Mamalahoa Hwy, HI 96743 Kamuela, USA; \\
\supit{f}Flat Wavefronts, 21 Lascelles Street, Christchurch 8022, New Zealand
}

\authorinfo{Further author information: M.L.: E-mail: masen@uvic.ca}

\begin{document} 
\maketitle 

\begin{abstract}
We propose and apply two methods to estimate pupil plane phase discontinuities for two realistic scenarios on VLT and Keck. The methods use both Phase Diversity and a form of image sharpening. For the case of VLT, we simulate the `low wind effect' (LWE) which is responsible for focal plane errors in the SPHERE system in low wind and good seeing conditions. We successfully estimate the simulated LWE using both methods, and show that they are complimentary to one another. We also demonstrate that single image Phase Diversity (also known as Phase Retrieval with diversity) is also capable of estimating the simulated LWE when using the natural de-focus on the SPHERE/DTTS imager. We demonstrate that Phase Diversity can estimate the LWE to within 30 nm RMS WFE, which is within the allowable tolerances to achieve a target SPHERE contrast of 10$^{-6}$. Finally, we simulate 153 nm RMS of piston errors on the mirror segments of Keck and produce NIRC2 images subject to these effects. We show that a single, diverse image with 1.5 waves (PV) of focus can be used to estimate this error to within 29 nm RMS WFE, and a perfect correction of our estimation would increase the Strehl ratio of a NIRC2 image by 12\%.

\end{abstract}



\section{Introduction}
\label{sec:intro}
\begin{figure}
    \begin{center}
    \includegraphics[scale=0.40]{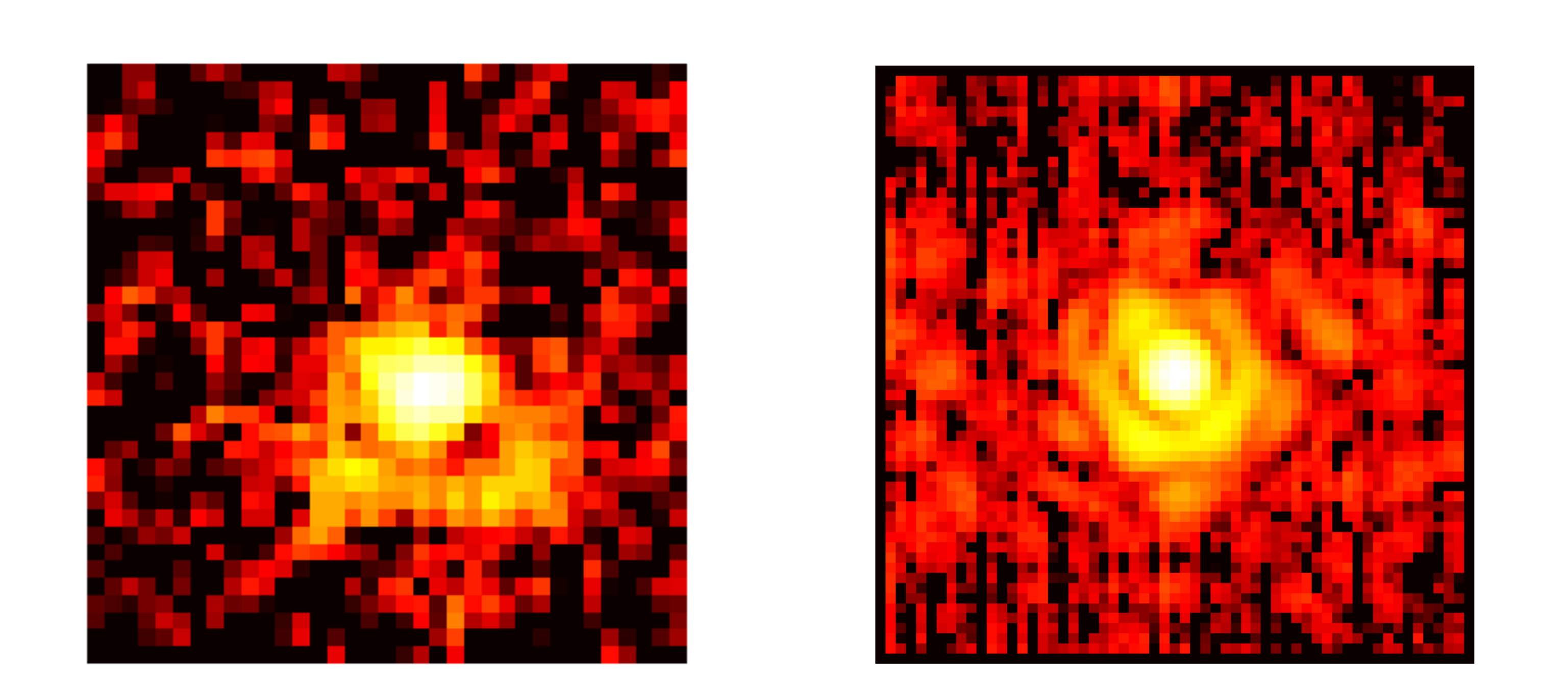}
    \vspace{1mm}
    \end{center}
    
    \caption{Left: An image acquired by the DTTS imager on SPHERE during a night with a strong low wind effect, shown in log-scale (courtesy of J.F. Sauvage). The asymmetric `ear' like features on the PSF shown here are an example of the PSF contamination experienced throughout the course of the entire night, and restrict the use of the instrument. Right: A K-band, short exposure Keck/NIRC2 image (also shown in log-scale) displaying typical features of `low order residuals', which are persistent throughout the duration of the closed AO loop (courtesy of S. Ragland).}

    \label{fig:exampleImages}
    \end{figure}

Piston discontinuities in segmented pupils are difficult to quantify when considering traditional adaptive optics (AO) systems, which originates from the inability of the Shack-Hartmann wavefront sensor (WFS) to estimate differential piston. A prime example of this is the low wind effect (LWE) on the VLT/SPHERE system, where nights with good seeing and a relative lack of wind surprisingly yield focal plane images of poor quality. This effect has been interpreted to be a result from temperature discontinuities across the VLT pupil\cite{Sauvage2015,Sauvage2016}. Conceptually these temperature discontinuities are thought to be defined by the secondary mirror spiders, which act as thermal barriers between segments. Airflow simulations of the spiders have been shown to reproduce these effects, and for more information we refer the reader to Sauvage et al. (2016)\cite{Sauvage2016}. Sharp differentials in temperature may translate to a differential piston effect, which is due to the index of refraction of air having a dependency on temperature. The Shack-Hartmann WFS is simply unable to detect differential piston and the resulting PSF reveals features akin to `Mickey Mouse Ears'\cite{Sauvage2015} (see Figure \ref{fig:exampleImages}); correcting this effect is crucial considering the purpose of SPHERE is to achieve the highest contrast possible, which is clearly contaminated by this effect. The target contrast of SPHERE is 10$^{-6}$ at 0.5"\cite{Beuzit2008}, which is achievable if the non-common path aberrations (NCPA) in the system are under 50 nm RMS WFE (wavefront error)\cite{Sauvage2016}. The internal NCPA of SPHERE were initially quantified at 25 nm RMS\cite{Sauvage2016b}, however this value is thought to have grown to $\sim$40 nm RMS and will be quantified in the near future (Fusco et al., in prep.). Assuming this latter value to be true, this leaves a maximum of only 30 nm RMS (the quadratic sum) from other contributions such as the LWE to achieve SPHERE's target contrast. The LWE is observed to occur when nights have sub meter per second speeds ($\sim$ 1/5 of the nights at Cerro Paranal) and therefore such a method to quantify and correct this effect could be very valuable for the proper functionality of SPHERE. This effect drives the operation of the observatory to disable the SPHERE instrument and re-organize the observations program. Developing the ability to estimate the LWE to within 30 nm RMS is therefore critical for the success of the SPHERE project.

Piston discontinuities also impact the performance of segmented telescopes such as the Keck telescopes. Artifacts of the PSF due to differential piston (and potentially other sources) on the Keck/NIRC2 AO system have been observed and identified as `low order residual errors' \cite{Rampy2014,Ragland2014} (see Figure \ref{fig:exampleImages}). These errors typically result in a PSF with a deviation in the first diffraction ring and a reduced Strehl ratio, and are shown to be stable on relatively long time scales (i.e. $>30$ minutes \cite{vanDam2016}). Some, but not all\cite{vanDam2016}, of these errors arise from the inability of the SHWFS to detect differential piston. Any correction of this effect requires a reliable estimation; applying such a correction will increase the Strehl ratio and allow the AO system to achieve its full potential. Estimating this effect will also provide essential information for PSF reconstruction, which is vital for a variety of astronomical applications\cite{Ragland2016}.

These two pupil effects have been investigated using a variety of different approaches. For example, recent studies (see N'Diaye et al. 2016\cite{NDiaye2016}) have explored the use of a Zernike WFS (ZELDA\cite{N'Diaye2013}), capable of detecting these piston variations on VLT/SPHERE. In the case of the Keck low order residuals, the approaches of both the Gerchberg-Saxton algorithm and Phase Diversity have been used to estimate these residuals both in simulation and on-sky\cite{vanDam2016,Ragland2016}.

In this paper we explore two different methods for estimating these pupil-discontinuity effects for simulated on-sky data. The first method is the well established approach of Phase Diversity, where known diverse images are compared to their synthetic counterpart to estimate the phase of an optical system\cite{Gonsalves82}. The second method is the technique of Focal Plane Sharpening\cite{Lamb2014} (FPS), where the PSF in the focal plane is optimized using only a deformable mirror. We simulate realistic images on both the VLT/SPHERE system and the Keck/NIRC2 system and investigate the feasibility of these methods to estimate the respective errors in question. We also explore the concept of single image Phase Diversity (also known as Phase Retrieval), which could be very useful in avoiding turbulence evolution and AO residuals between a set of on-sky images.

\section{Estimation methods}
\subsection{Phase Diversity}

 We employ a Phase Diversity code that follows the formulation of Paxman et al. (1992)\cite{Paxman92}, where an aberration-only objective function and its gradient are fed through a non-linear optimization algorithm to minimize the quadratic difference between synthetic and observed images. The aberration-only objective function consists of the coefficients of the basis to be estimated (i.e. Zernikes, or any other type of basis). The stopping criterion is defined by a tolerance parameter, which is described and determined in Lamb et al. (2016)\cite{Lamb2016_PD}; when the quadratic difference between the images is below this value convergence is reached. The synthetic and real data are typically in and out of focus images, although the code can incorporate any number of images or type of diversity. The optimization technique we employ is the well established quasi-Newton Broyden–Fletcher–Goldfarb–Shanno (BFGS) algorithm. The code has the ability to either jointly estimate the object along with the phase, or to just estimate the phase itself and assume the object known (which we simplify as a point source). This code has been developed as a class for the AO MATLAB software OOMAO\cite{OOMAO}; for more details of the code see Lamb et al. (2016)\cite{Lamb2016_PD}, and for an overview of the Phase Diversity technique we refer the reader to Mugnier et al. (2006)\cite{Mugnier-l-06a}. We adopt a 10\% error for each mode used in the creation of our diverse images to simulate realistic errors.
 
\subsection{Focal Plane Sharpening}

The results in this work use a MATLAB based Focal Plane Sharpening code, which has also been developed as a class for OOMAO\cite{OOMAO}, and has cross-compatibility with its Phase Diversity counterpart. The algorithm receives the focal plane PSF as input and optimizes on a variety of criteria (chosen by the user) using the Nelder-Mead downhill-simplex method\cite{Nelder-Mead}; the input parameters to the optimization method are the basis used to create the PSF (which can be DM influence functions, Zernike Modes, or any combination of modes chosen by the user). This method, along with a description of different criteria choices are explained in more detail in Lamb et al. (2016)\cite{Lamb2016_PD}. The results in this work use the following criteria: at each step in the optimization a small region centred on the PSF is extracted and subsequently median-filtered with a 2*FWHM kernel (i.e. two times the number of pixels across the FWHM of the theoretical diffraction limited PSF) to reduce noise; a 2D-Gaussian is fit to this image, from which the amplitude is measured. The magnitude of this amplitude is the metric that is optimized, changing at each iteration with the new set of basis coefficients.

\section{Estimating the Low Wind Effect on SPHERE}
\label{sec:LWE}
    \subsection{Basis and simulated images}
\begin{figure}[h]
    \centering
    \includegraphics[scale=0.55]{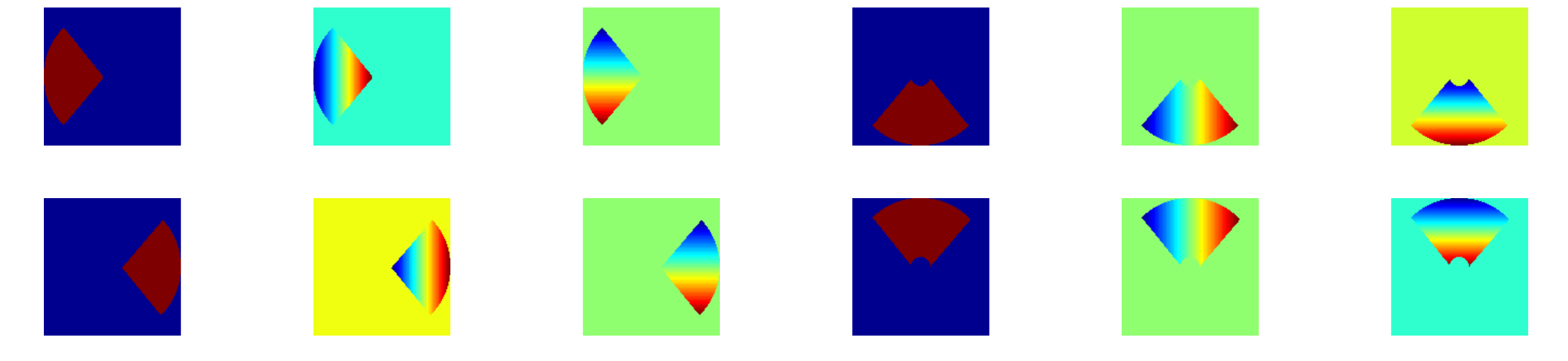}
    \caption{Piston, tip, and tilt basis used to recreate the PSF variations seen during the low wind effect on SPHERE. Each mode is normalized to 1 rad RMS (except the pistons). For the remainder of this paper, mode `1' of this basis corresponds to the top left mode shown here (piston on the left segment). The remaining modes numerically follow from left to right, ending with mode `12' shown in the bottom right of this figure (tip on the top segment).}
    \label{fig:vltBasis}
    \end{figure}

\begin{figure}[h]
    \centering
    \includegraphics[scale=0.5]{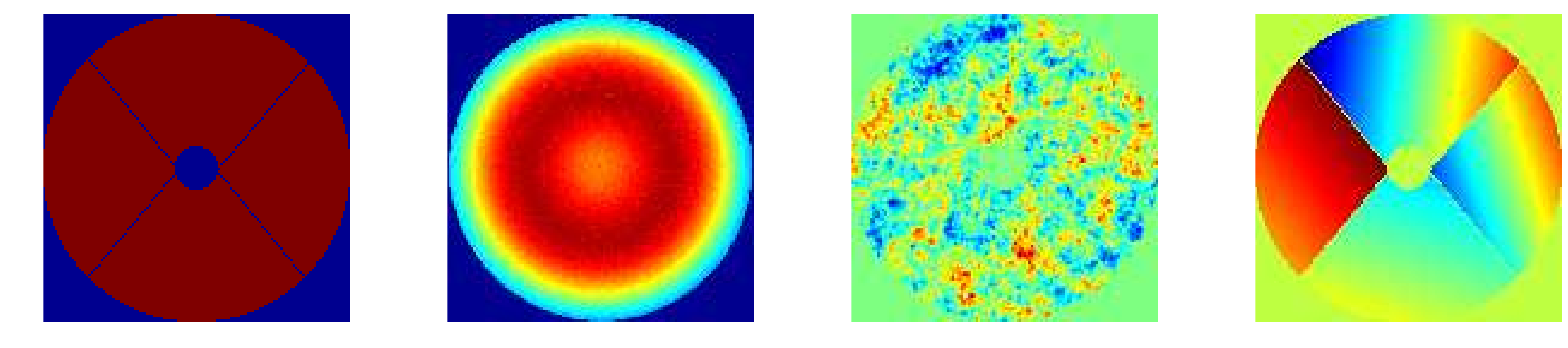}
    \caption{From left to right: VLT pupil, SPHERE apodization mask, assumed NCPA corresponding to 45 nm RMS WFE, and 1200 nm PV WFE low wind effect errors.}
    \label{fig:vltPupil}
    \end{figure}

    To estimate the low wind effect on SPHERE we propose a basis roughly defining the pupil plane phase variations that occur at each quadrant of the VLT pupil (see Figure \ref{fig:vltBasis}). The basic principle of our analysis is as follows: estimate the coefficients of this modal basis using both Phase Diversity and focal plane sharpening and assess the performance of each method. The amplitude of the piston errors we use to artificially produce the LWE are $\sim$ 1200 nm peak-to-valley (PV) WFE and are shown in Figure \ref{fig:vltPupil} (right); this LWE error is conservatively high, as typical LWE errors exist in the range of 600-800 nm PV WFE\cite{Sauvage2015}. The objective of our work is to estimate this error to within 30 nm RMS. We also adopt typical NCPA representive of the SPHERE system, specifically 45 nm RMS WFE following a 1/$\nu^2$ power law; they are also shown in the same Figure. As described in Section \ref{sec:intro}, the magnitude of the SPHERE NCPA are estimated to be $\sim$40 nm RMS, and our choice of 45 nm RMS is chosen to reflect a conservative over-estimate of the error. This over-estimate increases the error budget to 54 nm RMS, and we still aim to quantify the LWE to within 30 nm RMS WFE.
    
    Phase Diversity and FPS require focal plane images of the PSF in order to estimate the coefficients of this selected basis; these images are created as simulations of the SPHERE differential tip tilt sensor (DTTS) imager. The DTTS imager exists directly before the coronograph on SPHERE and therefore shares most of the common internal aberrations with the science image optics. Furthermore, using the DTTS images for the LWE characterization is much less complicated than using the coronographic images, even though the latter can be used for the estimation of aberrations\cite{Paul2014,Herscovici2017}. We therefore aim to show the DTTS imager is sufficient for the LWE estimation and create synthetic DTTS H-band images dominated by photon and read noise.
    
     The images are created using the adaptive optics simulation tool OOMAO with a 32x32 pupil sampling. Adding photon and read-noise is straightforward with this software, and we adopt a read-noise of 10e, as is typical for a Hawaii I detector. We simulate a DM with 41 actuators across the pupil. Images are created with a sampling as close to the DTTS imager as possible ($\sim$ 4 pixels across the FWHM). We subject the image to a turbulence profile (generated assuming an r$_0$ of 11 cm at 550 nm, and that all the turbulence occurs at the ground layer), and subsequently apply an AO correction using the simulated DM. The turbulence is evolved at a sampling rate of 500 Hz, and with a windspeed of 15 m/s; long exposure images are created by stepping the turbulence over the appropriate number of sampling steps pertaining to the total integration time of the image. This is particularly important since the DTTS imager typically acquires $\sim$1 second exposures. We note however that no residual AO phase errors are incorporated in the generation of these long exposure images (i.e. lag, aliasing, etc.) and we will have this long exposure functionality in the near future. However we are currently able to apply these residual phase errors to instantaneous images, and we generate these in a different analysis of Keck images in this paper. Finally, the images used with both Phase Diversity and FPS contain a field of view within the correctable region of the DM. This is not extremely important for focal plane sharpening, but for Phase Diversity it is extremely important: we find we have serious errors otherwise, increasing in effect as more of the uncorrected halo contaminates the image. Furthermore, the diversity we choose (i.e. focus) is always aimed to have its intensity distribution contained within the `dark' correctable region.
    
    \subsection{Estimation methods}
        \subsubsection{Phase Diversity}
        \label{ref:PDLWE}
        
        We first consider the estimation of the LWE modal coefficients by means of Phase Diversity; in particular we employ what we consider `Classic' Phase Diversity - when two images are used with focus diversity and the object is simultaneously estimated. Due to the combination of 45 nm RMS NCPA and the relatively large PV amplitude of the LWE ($\sim$1200 PV nm), we consider two waves PV of focus to ensure the diversity is larger than the phase to be estimated.

        It is important to simulate images with realistic DTTS signal-to-noise ratios (SNR), and so we consider the on sky data shown in Figure \ref{fig:DTTS}. We simulate a star with a typical SNR from this plot and run Phase Diversity with the aforementioned parameters. It was found that we have difficulty on convergence, where half the time the solution will `run away' in tip and tilt and converge on an erroneous result. We consider 3 approaches to solving this issue:
    
            \begin{figure}[h]
            \centering
            \includegraphics[scale=0.5]{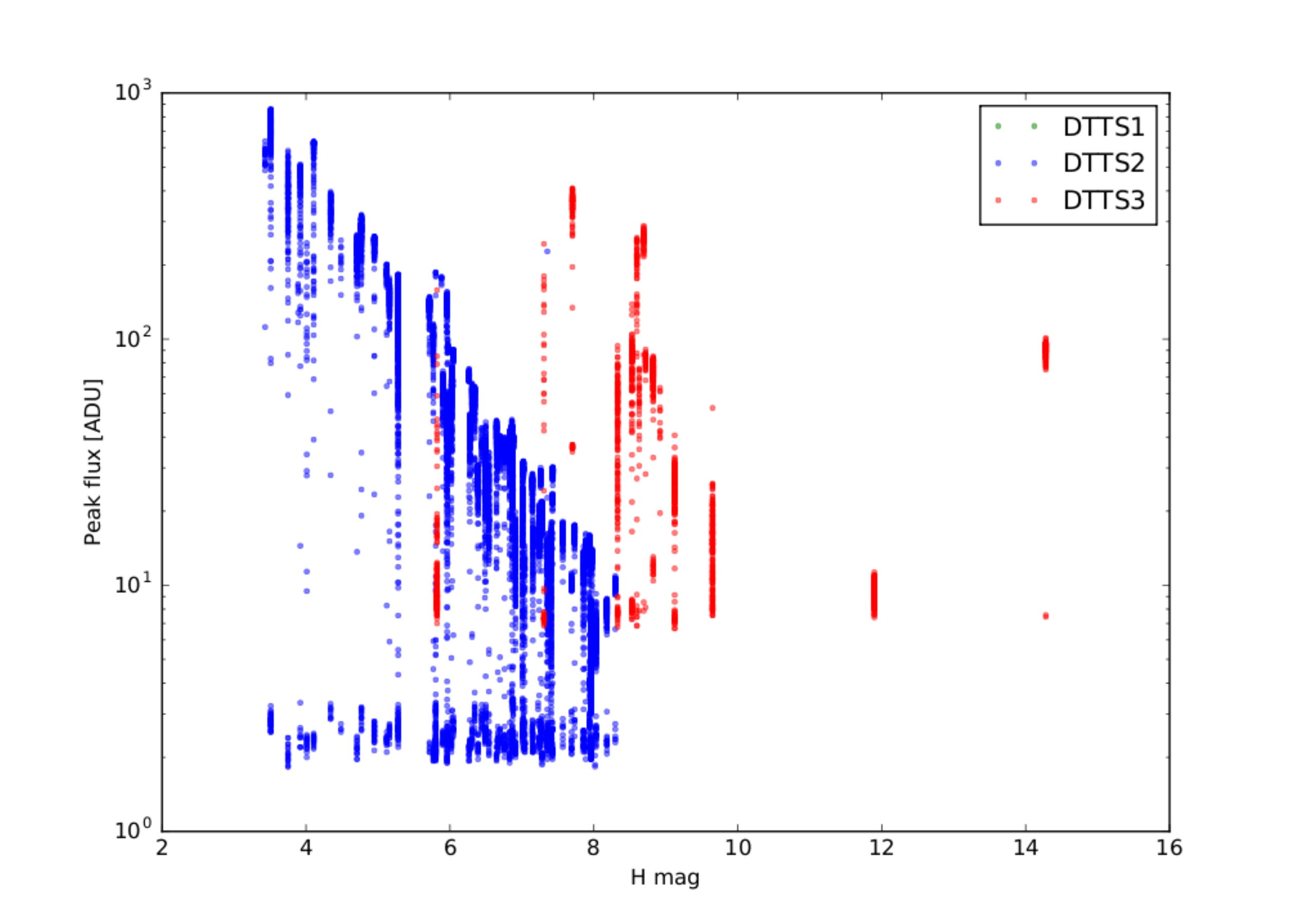}
            \caption{DTTS imager data, taken from SPHERE. The different colors correspond to different acquisition modes: blue points are taken in a mode optimized for bright stars while red points are suited for fainter stars; the green points represent an additional acquisition mode rarely used (and therefore explains the lack of points in this plot). The cloud of points around 2-3 ADU correspond to mis-detections, and we take this as the noise. Note: the values in this curve are subject to the inherent 20 nm PV focus on the DTTS imager, which results in a lower peak intensity than the true data shown here. After considering the noise floor and the data points adjusted for the 20 nm focus, we estimate a typical star has a SNR $\sim$ 70, and use this value for our analysis.}
            \label{fig:DTTS}
            \end{figure}   
    
\begin{figure}[h]
        \centering
        \includegraphics[scale=0.75]{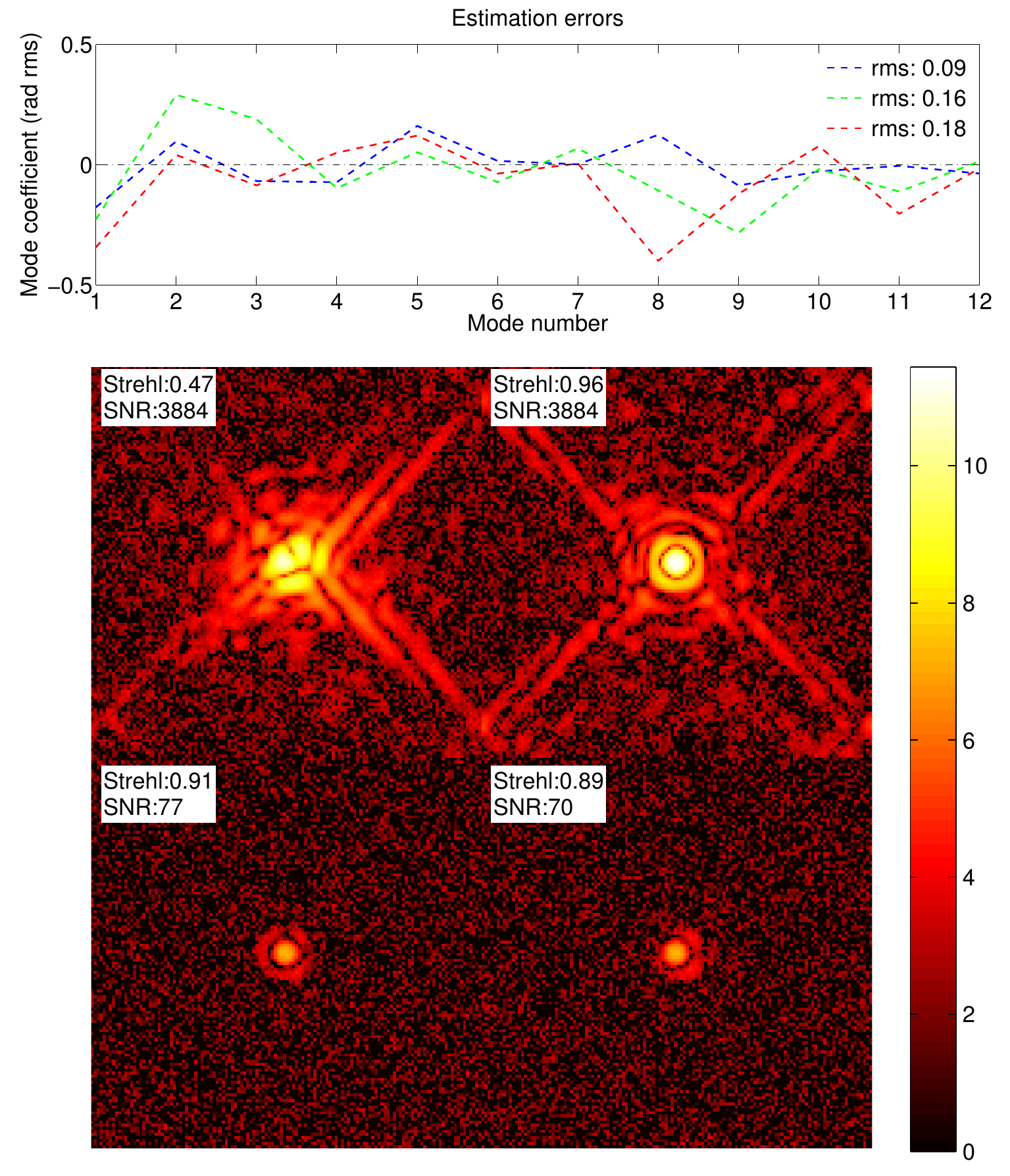}
        \caption{\textbf{Top:} residuals of LWE piston, tip, and tilt estimations from the actual modes, using Phase Diversity for 3 different scenarios (blue: long exposure object-estimation with focus diversity, green: object-estimation with higher diversity (Z$_{66}$), and red: assumed point source with focus-diversity). It can be seen the long exposure (blue) scenario performs the best, as indicated its RMS residual from the actual modes. \textbf{Bottom:} four panels of simulated VLT images, created from the phase projection of the estimated modes; they are described as follows: the upper two panels include no correction and long exposure Phase Diversity, respectively. The lower two panels include the higher diversity and assumed point source scenarios, respectively. From these images, the highest performance in terms of Strehl is clearly using the long exposure image. The bottom two images have diffraction rings that fall under the pedestal of the noise}
        \label{fig:LWE1}
        \end{figure}

            \begin{enumerate}
            \item Increase the SNR of the image (i.e. increasing the exposure time of the image), however on the real system this will integrate residual AO effects such as lag error, etc. We find this solution increases the Strehl from 0.47 to 0.96 (see Figure \ref{fig:LWE1}), however we do not simulate these long exposure errors and caution that these errors should be considered for a more in-depth analysis. Given the practical simplicity of this solution, it seems like a viable option. 
            
            \item The convergence seems to break down because of a signal-to-noise issue, therefore it is worth exploring how a different type of diverse image performs  under this same noise. Higher diversity modes, specifically cousins of the trefoil family (i.e. Z$_{11}$, Z$_{21}$, etc.), seem to work better in general in our simulation. In Particular we consider Z$_{66}$, where we have chosen this mode because it is a relatively high order mode - which we find in general produces better phase estimates, presumably because of the large diversity introduced to the PSF - and it is not too high such that it will be difficult to create with a SPHERE-like DM. Our simulation shows this mode always produces a better estimate of the phase (by a few nm RMS), and it converges faster than the focus-diversity case. We adopt 2 waves (PV) of Z$_{66}$ as our diversity and our simulation shows a Strehl increase from 0.47 to 0.91 (see Figure \ref{fig:LWE1} (bottom left image). This improvement is not as high as the previous example, however it is achieved with at a lower SNR. One problem may arise in realistically implementing this approach; high order phase speckles introduced from this diversity may be hard to disentangle from realistic AO phase residuals and high order, uncorrected NCPA. Furthermore, creating higher mode shapes such as Z$_{66}$ will always have a higher residual fitting error compared with a simpler mode such as focus. 
            
            \item Assume the object is known, which should be reasonable considering a star is effectively a point source, thus simplifying the Phase Diversity algorithm. The immediate result is that the estimate is not as accurate (acheiving a lower Strehl of 0.89 compared to the two aforementioned solutions), however it seems to have a much faster convergence rate. Figure \ref{fig:LWE1} shows results from our simulation using this method.
        \end{enumerate}
        
            From all three of these scenarios we conclude:
            \begin{itemize}
                \item Under typical DTTS imaging conditions, `Classic' Phase Diversity does not reliably work. If longer exposures do not contain too many adverse AO phase residuals, then Phase Diversity with long exposure images solves this issue. 
                \item Otherwise: using typical DTTS SNR values, Phase Diversity can work with a higher order diversity, or with assuming a known object. In the former case the estimate is more accurate, and in the latter the speed is considerably faster.
            \end{itemize}
            
            It is worth noting the diversity values adopted in this work were mainly chosen due to convention. In the context of Phase Diversity and Phase Retrieval, the exploration of induced diversity has been investigated by numerous examples in the literature. For example, Jurling et al. (2014)\cite{Jurling2014} consider scenarios with +/- 8 waves of focus diversity. Therefore, we acknowledge that it is possible alternative diversities may be more desirable, but for the purposes of this work we do not further pursue this topic.
    
    \subsubsection{Focal Plane Sharpening}
    \label{sec:LWEFPS}
    Given the fact that FPS is not model dependent, (unlike our form of Phase Diversity), it is worth investigating its performance under the same conditions as our Phase Diversity analysis. In general, our simulations suggest FPS works with even lower SNR images than Phase Diversity. However, one obvious limiting factor of FPS is the number of iterations (i.e. images) taken, particularly with the LWE considering the effect can vary over short time scales (i.e. tens of minutes). The LWE is observed to have time varying evolution over the course of several minutes (see Sauvage et al. 2015\cite{Sauvage2015}), and we arbitrarily choose a reasonable `window' in which to quantify this effect as one minute. Therefore, we implement the constraint that FPS should contain no more than 60 iterations, given the fact that a typical DDTS exposure is $\sim$ 1 second. In addition, we hypothesize that starting with an initial estimate from Phase Diversity may benefit from an improvement of FPS, based on its model independence. Furthermore, we hypothesize that the number of iterations (images) will be greatly reduced if using a starting point from Phase Diversity. Therefore we investigate 5 scenarios:
    
    \begin{itemize}
        \item Case-1: Phase Diversity on a typical SNR DTTS image, taking the fastest solution - which is when the object is assumed known (i.e. scenario 3 from the previous section).
        \item Case-2: Focal Plane Sharpening on the same type of image, starting from the null position.
        \item Case-3: Focal Plane Sharpening performed on the solution from the Phase Diversity example.
        \item Case-4: Focal Plane Sharpening performed on the best solution from the Phase Diversity example to explore if it does indeed outperform the model dependent Phase Diversity.
        \item Case-5: Focal Plane Sharpening starting from the lowest SNR image possible.
    \end{itemize}
    
    Table \ref{table:LWE} summarizes the results from the aforementioned cases. We note here that global tip and tilt are removed from the residual wavefront, where the residual wavefront is the known aberrated wavefront (LWE+NCPA) subtracted from that of the estimated wavefront. These residual wavefronts include the 45 nm RMS NCPA and will therefore be much larger than our target LWE estimation of 30 nm RMS. In case 4 and 5, FPS uses the PSF created from correcting the Phase Diversity estimate as input, in addition to using the Phase Diversity estimated modes as a starting position. It is found FPS improves the Phase Diversity result when the object is assumed known. However, there is no clear improvement from FPS on the best case Phase Diversity from Section \ref{ref:PDLWE} (where we used a longer exposure and estimated the object). In other words, in the case where the Phase Diversity images have a lower SNR, FPS will always improve those images. We find the FPS improvement takes $\sim$ 60 images when used with an initial Phase Diversity estimate, and for each image there is a very small computation time. Therefore we conclude this method would take between 1-2 minutes, which is slightly outside our prescribed time constraint of 1 minute.

    Interestingly, we also find that FPS run by itself will converge on a solution much different than the original modal coefficients (see the RMS residual coefficients column of Table \ref{table:LWE}), and furthermore this solution is of comparable residual WFE to our other best scenarios. The number of iterations for this convergence was about 120 - about double the number of images that used an intial estimate from Phase Diversity. Perhaps even more intriguing was the lower end SNR capabilities of FPS: it was found FPS could successfully converge on images with a SNR of 10, in about 120 iterations.
    
    Regardless, we find the best estimate of the LWE requires Phase Diversity with object estimation. For the remainder of Section 3, we further expand this method to assess its feasibility to accommodate SPHERE's WFE requirements as outlined in \ref{sec:intro}. In addition, we consider an alternative Phase Diversity scenario that requires only the raw DTTS images.
    
    \begin{table}
    \caption{Phase Diversity and Focal Plane Sharpening results correcting for the Low Wind Effect.}
    \hspace*{0.00cm}\begin{tabular}{@{}lcccccr@{}}
    \hline
    Case & Method 	                & Resid. coeff.**	    & Resid. wavefront*** 	& Strehl        & No. of    & Initial \\
         &	                & RMS (rad)	        & RMS (nm) 	        & (Marechal)    &     images          & SNR  \\
    \hline
Case 1 &    PD-1 (assumed object)   & 0.17	                & 89.4 	        & 89.1	        & 2             & 70 \\
       &    PD-2 (long exposure)*    & 0.09	                & 55.3 	        & 95.7	        & 2             & 3.8e4 \\
Case 2 &    FPS                     & 0.40	                & 56.1 	        & 95.5	        & 122           & 106\\
Case 3 &    FPS + PD-1              & 0.12	                & 57.9 	        & 95.3	        & 64+2          & 119\\
Case 4 &    FPS + PD-2              & 0.11	                & 55.7 	        & 95.6	        & 30+2          & 129\\
Case 5 &    FPS low SNR             & 0.28	                & 75.4 	        & 92.1	        & 121           & 10\\
    \hline 
    
    \label{table:LWE}
    \end{tabular}
    *Shown for reference\\
    ***The RMS of the difference between the estimated and actual coefficients\\
    ****The \textit{total} aberrated wavefront (LWE+NCPA) subtracted from the LWE estimate

    \end{table}

    \subsubsection{Estimating the LWE from a single image}
    \label{sec:singleImageLWE}

    Considering the evolution between two sequential images acquired in a closed loop AO system (due to changing seeing and AO residuals), it is desirable to consider Phase Diversity using only one image. Single image Phase Diversity is analogous to Phase Retrieval\cite{Gonsalves82}, and the image in question is subject to diversity; for the rest of this paper we will refer to this technique as `single image Phase Diversity' instead of Phase Retrieval to stay consistent with nomenclature of the multiple image scenarios. Work has been done on-sky in the past showing the challenges involved with using two sequential Phase Diverse images\cite{Jolissaint2012}. Furthermore, the DTTS has a natural focus amplitude of 20 nm RMS, providing a diverse image with no reference slope manipulation. We have found that when using a non-simple pupil - such as the case with the apodized VLT pupil considered here - that single image Phase Diversity works for this amplitude of natural focus. This approach assumes the object is known, in which case we assume a point source (which is not unreasonable considering we are simulating stars). In Lamb et al. 2016\cite{Lamb2016_PD}, we consider the limitations of single image Phase Diversity (i.e. for uniform, circular, symmetric pupils). However, for this paper we will not explore the technical background of this technique. It is worth noting similar work has been done investigating single image Phase Diversity\cite{Meimon-a-10a} in developing the LIFT technique, and it is not an entirely new concept. Furthermore, the concept of a non-simple pupil to facilitate phase estimation dates as far back as 1965 by C.L. Mehta\cite{Mehta1965}.

    To validate the performance of single image Phase Diversity we consider 3 different scenarios of LWE estimation: 
    \begin{itemize}
        \item i) `Classic Phase Diversity', in which case phase and object are simultaneously estimated from two images with 0 and +2 waves PV of focus.
        \item ii) Two image Phase Diversity with \textit{no} object estimation, using images with -1 and +2 waves PV of focus.
        \item iii) One image Phase Diversity, using a single image subject to 20 nm PV focus, similar to the actual DTTS images.
    \end{itemize}

\begin{figure}

    \begin{center}
    \includegraphics[scale=0.55]{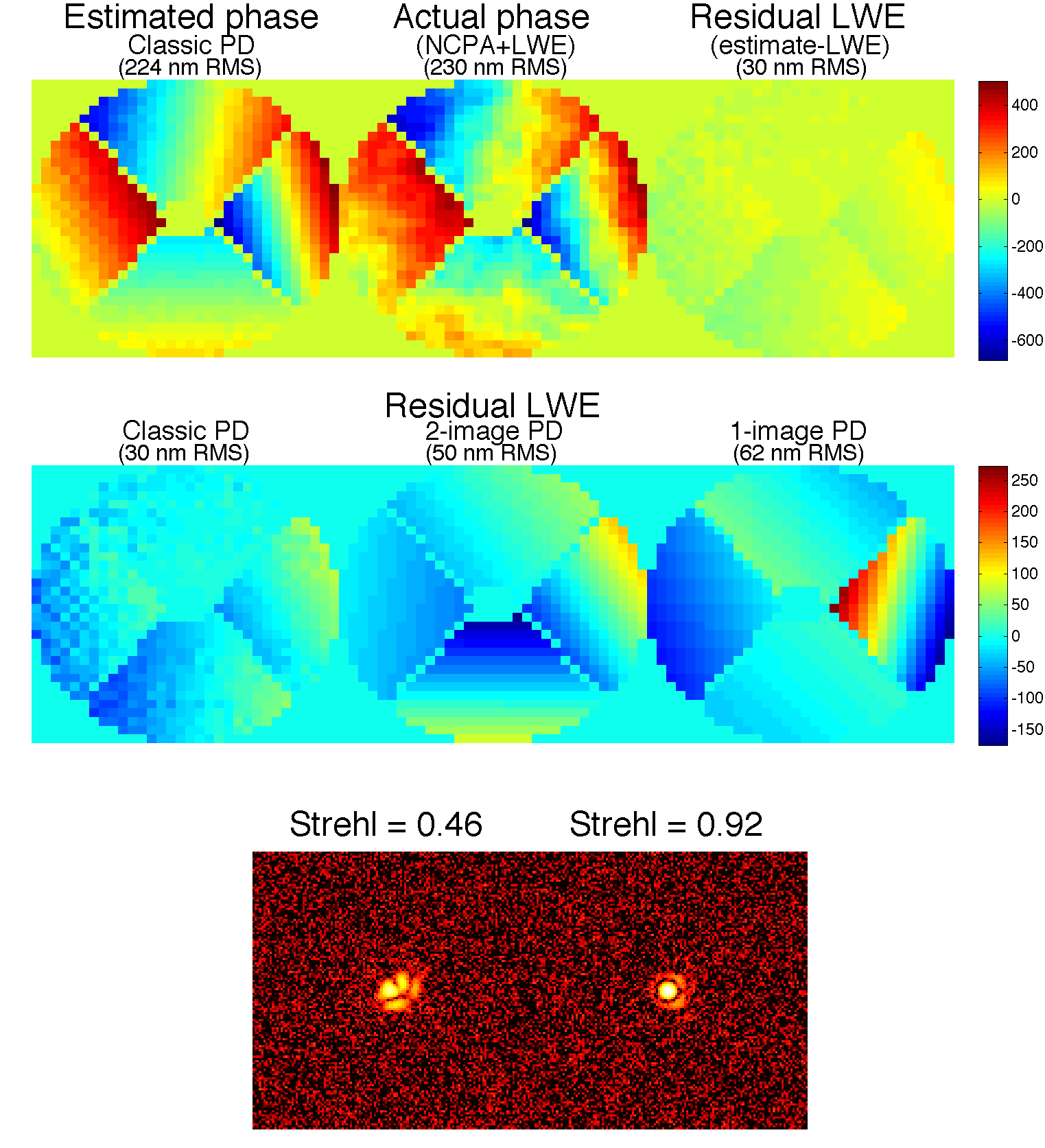}
    \end{center}

    \caption{\textbf{Top:} Estimated LWE (left) from Classic Phase Diversity (phase and object estimation using images with 0 and +2 waves PV focus), actual phase injected (center), and residual phase between the estimate and actual LWE injection (right). The residual WFE reaches the desired 30 nm RMS, such that a perfect correction of this estimated phase would result in a contrast at least 10$^{-6}$. \textbf{Middle:} Residual phase maps for 2 image Phase Diversity with and without object estimation (left and center panels, respectively) and single image Phase Diversity (right panel, using a single image with the natural focus of the DTTS imager). These additional Phase Diversity scenarios do not meet the performance of Classic Phase Diversity, but are shown here for comparison. The case of the single image should be considered useful for its potential of both a quick LWE quantification and unobtrusiveness in image acquisition. \textbf{Bottom:} Simulated PSFs before and after (perfect) correction from the single image LWE estimate.}
    \label{fig:singleLWE}
    \end{figure}

    The images used in the above scenarios have a SNR of $\sim$ 70 for an in-focus image; this is to stay consistent with Section \ref{ref:PDLWE}. In the case of Classic Phase Diversity, however, the solution did not converge with this value and a minimum SNR of $\sim$ 700 was required. Figure \ref{fig:singleLWE} summarizes the results of each of these scenarios. The best estimation is achieved by Classic Phase Diversity; the residual between the phase estimate of the LWE and the actual injected LWE is 30 nm RMS (top right of the Figure). However, in the cases of ii) and iii) LWE estimations are marginally worse, with 50 and 62 nm RMS residuals (respectively). Two image Phase Diversity with no object estimation is considered here strictly as a comparison with single image Phase Diversity, where the only real difference between the two scenarios is a single image. From these results it appears single image Phase Diversity with the natural DTTS focus can reasonably estimate the LWE, although not at the same performance of Classic Phase Diversity.
    
    \subsection{Performance evaluation}
    \label{sec:LWE_eval}
    Through consideration of both FPS and Phase Diversity, it appears the most viable form of LWE estimation (in terms of both speed and accuracy) is in some form of Phase Diversity. As mentioned in Section \ref{sec:intro}, 30 nm RMS WFE of the LWE is small enough to still achieve the target contrast of SPHERE and therefore a perfect correction of our from a Classic Phase Diversity estimate would achieve this requirement. We note here that the LWE in this simulation is representative of a larger-than-normal LWE night, therefore we expect better performance under less severe conditions. That being said, Classic Phase Diversity requires DTTS images with a higher SNR than those delivered by the operation of the sensor, and therefore would be subject to atmospheric residuals over the course of long exposures. Furthermore, the acquisition of diverse images would require reference slope manipulation; it is worth considering single image Phase Diversity requires no such manipulation when using the natural DTTS focus. Furthermore, the results shown here suggest the impact in LWE estimation is not terribly drastic between single image and Classic Phase Diversity (i.e. on the order of 5\% Strehl loss using the Marechal approximation), and could be seriously considered for estimation/correction if the science does not require the full 10$^{-6}$ contrast. Lamb et al. (in prep) are currently analyzing a sequence of DTTS images recently acquired at the VLT during a night subject to a strong LWE in order to better understand the feasibility of single image Phase Diversity.
    
\section{Estimating the Segment Piston Errors on Keck}

As previously mentioned, significant low-order AO residuals in the Keck/NIRC2 system exist and may originate in the form of co-phasing errors of the primary mirror segments. We now consider estimating these segment-piston errors employing the same single image Phase Diversity approach that was used in quantifying the LWE in Section \ref{sec:LWE_eval}. The motivation for this approach is the desire to avoid unwanted evolutionary effects between two sequential images, as has been discovered with Classic Phase Diversity in the past (as previously mentioned in \ref{sec:singleImageLWE}). We note that previous work has aimed at mitigating these effects by considering long exposure images with Phase Diversity, which average out the AO residuals and seeing effects \cite{Mugnier-a-08}, however in this work we are interested in assessing the feasibility of using short exposure images. Since we are interested in avoiding evolutionary effects between sequential images, we do not consider Focal Plane Sharpening in this exercise.

\begin{figure}
\centering
\includegraphics[scale=0.5]{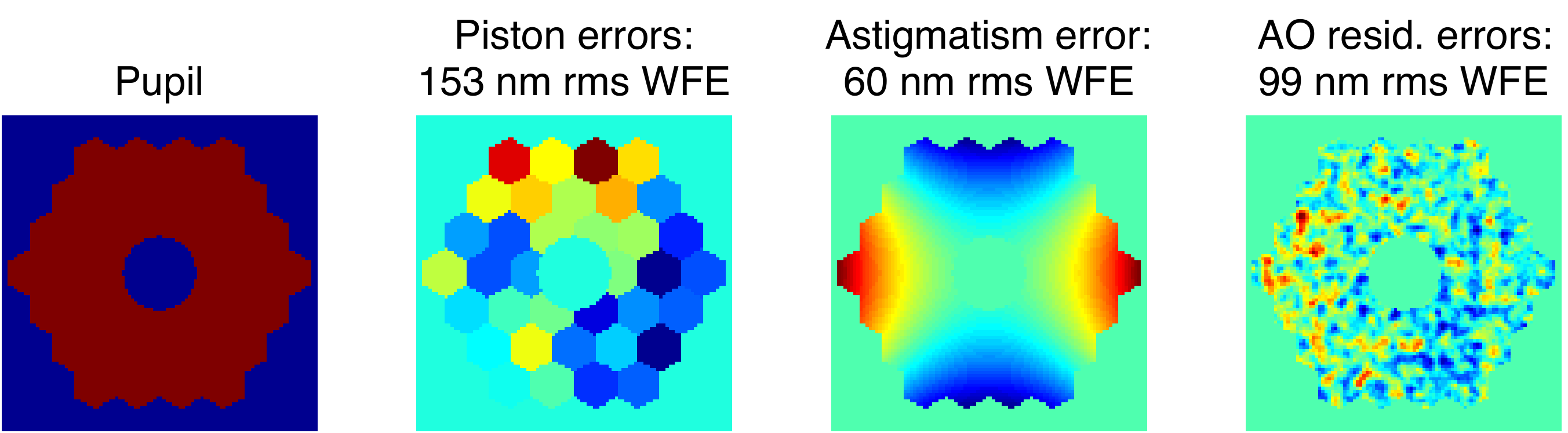}

\caption{From left to right: Keck pupil, simulated piston phasing errors to be estimated, inherent astigmatism of NIRC2, AO phase errors (i.e. servo lag, aliasing, photon noise, fitting). The three phase maps on the right are used to create our simulated Keck images.}
\label{fig:keckPhase}
\end{figure}

A defocussed NIRC2 image, subject to 153 nm RMS WFE co-phasing errors, is simulated as faithfully as possible. The co-phasing error injection we adopt here was taken from a similar phasing-residual analysis (using different estimation algorithms) \cite{vanDam2016}; the co-phasing error phase map is shown in Figure \ref{fig:keckPhase}. The simulated images were created on a pupil with 32x32 pixel sampling, and were subject to photon and read-noise errors; we consider a read-noise of 60e- (a single read-out NIRC2 image is closer to 40e-\footnote{http://www2.keck.hawaii.edu/inst/nirc2/Manual/ObserversManual.html\#Section2.2}, however we choose a larger value to be conservative). The images are generated with OOMAO in the environment of a simulated AO system with 21 actuators across the pupil (identical to the actual NIRC2 system). In addition to the co-phasing residuals, we also inject 60 nm RMS of astigmatism (a rough representation of the true non common path error), and 99 nm RMS of simulated AO residual errors (summarizing contributions from servo-lag, aliasing, and dm-fitting errors). Figure \ref{fig:keckPhase} shows phase maps representing these errors.

\begin{figure}
    
    \includegraphics[scale=0.58]{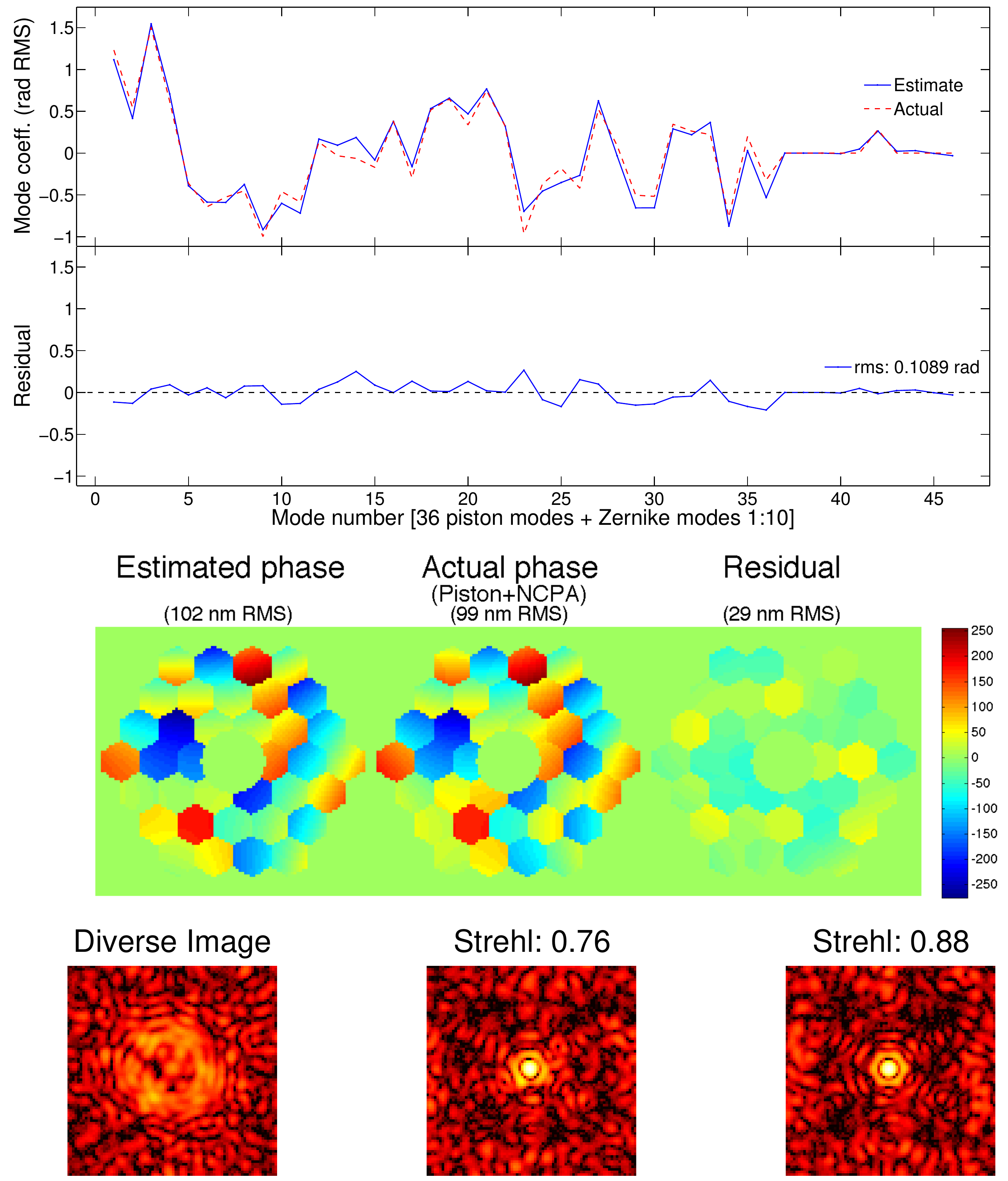}
    \caption{Estimation of both segment piston errors and NCPA (astigmatism) from a single diverse image of a simulated bright star (diverse image shown in bottom left). \textbf{Top:} Estimated modal coefficients of 36 piston modes and 10 Zernike modes (Z$_1$-Z$_{10}$). \textbf{Middle:} Phase reconstructed from the estimated modes (left) compared with the actual piston plus astigmatism phase (middle); the residual between the two is shown on the right with a WFE of 29 nm RMS. The global tip and tilt was removed from the estimated and actual phases, reducing the original co-phasing error from 153 to 117 nm RMS. \textbf{Bottom:} Simulated images of the initially aberrated system (middle) and situation where the perfect correction of the estimated phase is achieved (right).}

    \label{fig:KeckNCPA}
    \end{figure}

Assuming the object is a point source, the de-focussed NIRC2 image is estimated with our Phase Diversity algorithm. The amount of defocus is 1.5 waves (PV), and was chosen because any larger values caused a focus pattern that crept into the uncorrected halo. The estimated basis modes are pure piston variations, defined by each mirror segment (therefore a 36 element basis), along with the first 10 Zernike modes. One of the goals of our estimation is to see if it can decouple the piston errors from the astigmatism while under the influence of realistic AO residuals. Figure \ref{fig:KeckNCPA} summarizes our findings. The top of the figure shows the coefficients of the estimated piston plus Zernike basis in blue, overlaid on the actual modes injected into the system (red). The RMS residual difference between the actual and estimated modes is 0.1092 rad, which translates to 29 nm RMS. To visualize this estimation in terms of phase errors, the middle section of the figure displays the estimated phase (left), the actual NCPA plus segment error injection (middle), and residual phase of 29 nm RMS (right). Here the global tip and tilt is removed from both the estimated and actual phase prior to the subtraction. Hence the magnitude of the original 153 nm RMS co-phasing error has been reduced to 117 nm RMS; when this error is added to the 60 nm RMS of NCPA the final WFE is 99 nm RMS.

\subsection{Performance evaluation}
To get a sense of the performance of this analysis, we consider the situation where a perfect correction of the phase estimation can be achieved. The bottom of Figure \ref{fig:KeckNCPA} displays an image before and after such a correction, showing a Strehl improvement of $\sim$ 12\% (calculated directly from the images, using the ratio between the image OTFs and their diffraction limited counterpart). The signal to noise of a faint source increases proportionally to the Strehl ratio, therefore the exposure time is proportional to the square of the Strehl. In our example, assuming a perfect correction of the segment-piston estimation, the exposure time of a faint object would be $\sim$0.75 times less to achieve the same signal to noise without correction.
    
\section{Conclusion}

Segmented pupil error phase discontinuities were successfully estimated for two realistic scenarios on both VLT and Keck. From this work we conclude:

\begin{itemize}
    \item For SPHERE: typical DTTS imaging conditions make it difficult for `Classic' Phase Diversity, and we propose 3 solutions that improve the estimation, which we show in simulation. The best estimation was achieved by taking longer exposure images than typical of the DTTS, in addition to object estimation. However, our simulation does not model the long exposure AO residual effects such as servo-lag, aliasing and fitting error, which would provide additional uncertainties in a true estimation. 
    If in reality these residuals have a larger than desired impact on the estimation, we propose a workaround where a suitable LWE estimation is achieved by either using a higher diversity mode, or leaving out the object estimation (and assuming a point-source). The latter method is less computationally expensive with respect to any phase plus object estimation.
    \item If using the non-object estimation method described above, then Focal Plane Sharpening will improve the image (on top of the Phase Diversity estimation) in around 60 iterations, which is around 1-2 minutes assuming each image takes $\sim$1 second. However, if starting from the null position, Focal Plane Sharpening can take more than twice this number of iterations. Worthy of note, however, is that Focal Plane Sharpening can work on extremely low SNR ($\sim$ 10) DTTS images. 
    \item Running low SNR FPS on top of low SNR Phase Diversity yields roughly the same estimate as the best case Phase Diversity estimation, where a high SNR (i.e. long exposure) is required.
    \item We find that single image Phase Diversity is a useful tool to estimate the LWE. If using a single image with the natural focus on the DTTS imager, we find that this technique can estimate the LWE to almost the same degree as two images with no object estimation (using +1 and -2 waves of focus). However, two images \textit{with} object estimation significantly improves the phase estimation, albeit higher SNR on the images is required. The single image approach could be useful considering sequential on-sky images will have evolution from one image to the next - an effect known to cause issues with Phase Diversity in the past \cite{Jolissaint2012}. Lamb et al. (in prep) are currently analyzing on-sky DTTS data taken during a LWE sequence to further explore the feasibility of single image Phase Diversity.
    \item To reach the target contrast of 10$^{-6}$, the LWE must be estimated to within an error of 30 nm RMS, as discussed in Section \ref{sec:intro}. We find this goal can be achieved when using `Classic' Phase Diversity (an in/out of focus image with object estimation), given the SNR of the in-focus image is $\sim$700. The error on this estimation roughly doubles when we use Phase Diversity with a single DTTS image with a natural 20 nm RMS of focus, although on an image with a SNR $70$. We conclude that at a small sacrifice in Strehl ($\sim4\%$) and therefore contrast, single DTTS images with no manipulation could be used to quantify the LWE.
    \item The correction of the LWE effect using the wavefront control capabilities of SPHERE is a complex point, and not studied in this paper. The SPHERE system will benefit from a change of the spider coating, which will largely decrease the amplitude of the effect. Ideally the addition of this coating, combined with some sort of wavefront control scheme from the LWE estimation will be able to correct for the total effect in order to achieve the design contrast of SPHERE. Some recent tests have been investigated to prove that SPHERE wavefront control is able to produce the corresponding amplitude estimates of LWE (Sauvage et al. in prep.).
    
    \item Phase segment piston errors were successfully estimated for Keck/NIRC2 simulated images given the assumption that images can be acquired near-instantaneous (i.e. with `frozen' turbulence). This assumption should not be too unreasonable considering the limitation is only on the shutter speed of the detector (the availability of bright stars should not be an issue). A goal of this exercise was to show the capability of using Phase Diversity with a single, diverse image; this technique avoids any significant evolution effects between sequential images and would be a useful comparison to the complementary approach of using long exposure images for phase estimation \cite{Mugnier-l-06a}. Given an aberrated image subject to 153 nm RMS WFE of simulated co-phasing errors, we find we can estimate the phase to within an error of 29 nm RMS. Assuming a perfect application of this estimate can be applied to the mirror segments, this would result in a Strehl increase of 12\% for our simulated NIRC2 images. A direct result of such a fix would decrease the exposure time on faint sources by a factor of 1.3 to achieve the same SNR if no fix was applied.
   
\end{itemize}

\acknowledgments     
The research leading to these results received the support of the
A*MIDEX project (no. ANR-11-IDEX-0001-02) funded by the ”Investissements
d'Avenir” French Government program, managed by the French National
Research Agency (ANR). This research was also funded in part by a MITACS/Campus France Globalink Research Award (ref: IT06712).

All the simulations and analysis done with the object-oriented MATLAB AO
simulator (OOMAO) freely available from
https://github.com/cmcorreia/LAM-Public

\bibliography{report}   
\bibliographystyle{spiebib}   

\section*{Authors' Biographies}
\textbf{Masen Lamb} is currently a PhD candidate in the department of Physics and Astronomy at the University of Victoria, Canada. He received his BSc in Astronomy at the University of British Columbia, Canada. His current research involves the use of adaptive optics (AO) to resolve metal-poor stars in dense populations such as the Galactic Centre. His research also focuses on the calibration of non-common path aberrations in contemporary AO systems.

*The remaining Authors' biographies are unavailable.

\listoffigures
\end{document}